\documentclass[preprint,prb,showpacs,superscriptaddress]{revtex4}

\usepackage{graphicx}
\usepackage{amsmath}

\newcommand{\beq}{\begin{equation}}
\newcommand{\eeq}{\end{equation}}
\newcommand{\beqa}{\begin{eqnarray}}
\newcommand{\eeqa}{\end{eqnarray}}
\newcommand{\vep}{\varepsilon}

\begin{document}


\title{ Tunneling spectroscopy of spin-selective Aharonov-Bohm oscillations
        in a lateral triple quantum dot molecule }

\author{Y.-P. Shim}
\affiliation{Quantum Theory Group, Institute for Microstructural Sciences,
             National Research Council of Canada,
             Ottawa, Canada K1A 0R6}

\author{F. Delgado}
\affiliation{Quantum Theory Group, Institute for Microstructural Sciences,
             National Research Council of Canada,
             Ottawa, Canada K1A 0R6}
\affiliation{Departamento de Fisica Aplicada, Universidad de Alicante,
             San Vicente del Raspeig, 03690, Spain}

\author{P. Hawrylak}
\affiliation{Quantum Theory Group, Institute for Microstructural Sciences,
             National Research Council of Canada,
             Ottawa, Canada K1A 0R6}

\date{\today}

\begin{abstract}
We present a theory of tunneling spectroscopy of spin-selective
Aharonov-Bohm oscillations in a lateral triple quantum dot molecule.
The theory combines exact treatment of an isolated many-body system
with the rate equation approach when the quantum dot molecule is weakly connected
to the leads subject to arbitrary source-drain bias.
The tunneling spectroscopy of the many-body complex is analyzed
using the spectral functions of the system and applied to holes in a quantum dot molecule.
Negative differential conductance is predicted and explained as a result of
the redistribution of the spectral weight between transport channels.
It is shown that different interference effects on singlet and triplet hole states
in a magnetic field lead to spin-selective Aharonov-Bohm oscillations.
\end{abstract}

\maketitle


\section{Introduction}

Quantum dot (QD) systems are artificial quantum mechanical systems
with great controllability of the electronic, spin, and transport properties.\cite{jacak_hawrylak_book1998}
The occupancy and spins of electrons in single and coupled semiconductor QD systems can be controlled
electrostatically.\cite{ciorga_sachrajda_prb2000,koppens_buizert_nature2006,petta_johnson_science2005,%
vidan_westervelt_jsupercond2005,gaudreau_studenikin_prl2006,ihn_sigrist_njphys2007}
In lateral structures where QDs are defined by lateral gates on top of
a heterojunction containing two-dimensional electron gas (2DEG),
the confinement potentials and tunneling barriers can be easily tuned by applying voltages to the gates.
Coupled QD systems, where the coherence of different quantum states can play an important part,
are one of promising candidates for realizing
quantum information and computation devices.\cite{awschalom_loss_snt2002,%
brum_hawrylak_sm1997,loss_divincenzo_pra1998,sachrajda_hawrylak_book2003}
Coulomb interactions in such small systems become more important and
QD networks offer great flexibility for studying
many-body effects in a very controllable manner.
Transport through a strongly interacting QD system leads
to many interesting phenomena such as Coulomb blockade \cite{jacak_hawrylak_book1998}
due to the strong Coulomb interaction
and spin blockade in a double QD system with an electron localized
in one of the dots.\cite{ono_austing_science2002,johnson_petta_prb2005}
High source-drain bias transport can be used as a spectroscopic tool
for the interacting many-body system.\cite{pfannkuche_ulloa_prl1995,hawrylak_lectnotphys1997}

\begin{figure}
\includegraphics[width=0.8\linewidth]{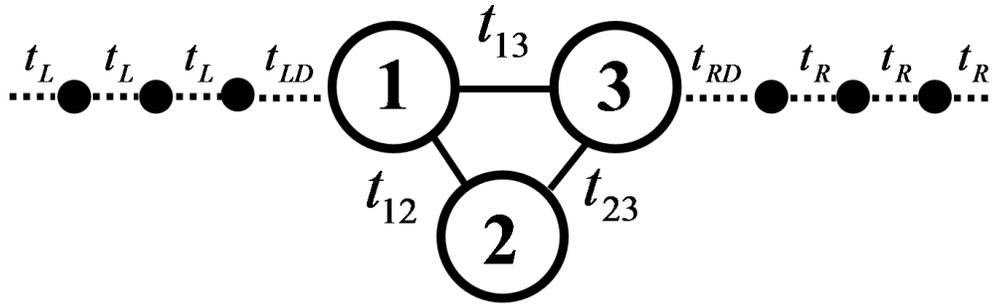}\\
\caption{Schematic diagram of a triple quantum dot molecule connected to leads.
         $t$'s are the tunneling matrix elements that connect two different sites.
         Each lead is modeled by a semi-infinite noninteracting chain
         which is connected to the TQD at one end.}
\label{fig:system}
\end{figure}

In triple quantum dot (TQD) systems,
the electronic and spin properties depend on the topology of the system.\cite{korkusinski_gimenez_prb2007}
Interference effects are expected in triangular TQD molecules\cite{delgado_shim_prb2007}
and coherent transport was experimentally observed.\cite{gaudreau_sachrajda_icps2006,ihn_sigrist_njphys2007}
We study a TQD system which is connected to two leads [Fig.~\ref{fig:system}],
with dot 1 connected to the left lead and dot 3 connected to the right lead.
In linear response regime, for an electron to move from one lead
to the other we need three different charge configurations
$(N_1,N_2,N_3)$, $(N_1+1,N_2,N_3)$, and $(N_1,N_2,N_3+1)$
in terms of the number of electrons in each dot to be on resonance.
That is, $E(N_1,N_2,N_3)+\mu_0 = E(N_1+1,N_2,N_3) = E(N_1,N_2,N_3+1)$
where $\mu_0$ is the chemical potential of the leads.
This allows an electron to move from the left lead to dot 1, to dot 3, and finally to the right lead.
If dots are not on resonance, a finite bias must compensate for the deviation from the resonance.
If another charge configuration $(N_1,N_2+1,N_3)$ with the extra electron in dot 2 is also on resonance,
we have four degenerate charge configurations and we define this point as a quadruple point (QP).
If the constituent QDs are strongly connected, we can not use simple classical arguments for the transport
and QP is defined as a point in parameter space where four different charge configurations have the same probabilities.
At QPs, the extra electron can move along two alternative paths before it escapes
and this leads to the Aharonov-Bohm (AB) oscillations \cite{aharonov_bohm_pr1959}
in the presence of a perpendicular magnetic field.
With empty systems, we can use single-particle picture for the transport and
a theoretical calculation for empty TQD using transfer matrix
was recently reported.\cite{delgado_hawrylak_jpcm2008}
For the transport involving states with more than one electron,
the strong Coulomb interaction and the correlation
between different charge configurations must be taken into account.
A rate equation for the probabilities of each quantum many-body state
of the system has been used to describe the transport
through strongly interacting quantum systems
weakly connected to the leads.\cite{averin_korotkov_prb1991,beenakker_prb1991,%
schoeller_schon_prb1994,weymann_konig_prb2005,%
bonet_deshmukh_prb2002,turek_matveev_prb2002,mitra_aleiner_prb2004,koch_oppen_prb2004,muralidharan_datta_prb2007}
In our previous work \cite{delgado_shim_prl2008},
it was shown that the interplay between the AB oscillations and spin blockade
in a TQD around a QP with an electron trapped in dot 2
leads to spin-selective AB oscillations.
Around this QP, the singlet states show strong AB oscillations
while AB oscillations for triplet states are suppressed.
Combined with the Zeeman splitting, these different behaviors of singlet and triplet states
result in repeated peaks of spin-down current at lower magnetic fields
and large spin-up current at high magnetic fields.

In the current work, we describe in detail how we calculate
the transport through a strongly correlated many-body system
based on rate equation approach in the sequential tunneling regime.
Higher order processes such as cotunneling will be
neglected.\cite{ingersent_ludwig_prl2005,kuzmenko_kikoin_prl2006,zitko_bonca_prb2008}
We will focus on a TQD system which contains one or two holes.
A hole is defined as absence of electron with respect to the fully occupied TQD with six electrons.
In a resonant TQD where all three dots are identical,
the two-hole system has spin-triplet ground state
while two-electron system has spin-singlet ground state.\cite{korkusinski_gimenez_prb2007,delgado_shim_prb2007}
The singlet-triplet gap is proportional to the inter-dot tunneling.
In other words, the inter-dot tunneling in triangular TQD
favors spin-singlet for two-electron system and favors spin-triplet for two-hole system.
Two-hole system in TQD has a very interesting property
that the spin structure of the system can be tuned by gate voltages,
which is in contrast to the two-electron system where the ground state is always spin-singlet.
By lowering the level energy of dot 2 (or raising the hole level energy of dot 2),
two holes are localized in dot 1 and 3, and
the ground state changes from triplet to singlet.\cite{shim_hawrylak_prb2008}
On the other hand, when we lower the hole level of dot 2,
we can permanently trap one hole in dot 2 with the additional hole moving around the TQD.
The four charge configurations with a trapped hole in dot 2 are
(1,1,2),(2,0,2),(1,2,1), and (2,1,2) in terms of electron occupation numbers.
As is shown later, at the QP where these configurations are degenerate,
spin singlet and triplet states of two-hole system are degenerate.
This degeneracy between singlet and triplet states implies
that the spin state is very sensitive to the external environment such as nuclear spin,
and it could be used for, {\it e.g.}, probing nuclear spins.
We investigate the transport around this QP
in the presence of a perpendicular magnetic field.

The plan of the paper is as follows.
In Sec. \ref{sec:Method}, we describe our model Hamiltonian
for the TQD, the leads, and the coupling between the TQD and the leads.
The current is expressed using electron and hole spectral functions
and master equation for the probabilities of each state of the system will be given.
In Sec. \ref{sec:Transport}, general conditions for transport
through a quantum many-body system in the sequential tunneling regime are discussed
in terms of transport channels, active state and trap state.
In Sec. \ref{sec:Results}, using the methods developed in Sec. \ref{sec:Method},
numerical results in a QP with a trapped hole are presented.
Negative differential conductance and spin-selective AB oscillations
are predicted and discussed in detail.
A brief summary will be given in Sec. \ref{sec:Conclusions}.

\section{Method}\label{sec:Method}

\subsection{Model}

We have shown previously\cite{korkusinski_gimenez_prb2007,delgado_shim_prb2007}
that the electronic properties of the TQD molecule with few confined electrons
($N=1$ to $6$) can be understood in the frame of the Hubbard model with one orbital per dot.
The Hamiltonian of the TQD subject to a uniform perpendicular magnetic field,
${\bf B}=B {\bf \hat{z}}$, is given by
\begin{eqnarray}
\widehat{H}_{TQD}
&=& \sum_{i=1}^{3}\sum_{\sigma} \vep_{i\sigma} d_{i\sigma}^{\dag} d_{i\sigma}
+ \sum_{i\neq j}\sum_{\sigma} t_{ij}(B) d_{i\sigma}^{\dag} d_{j\sigma}
+ \sum_{i} U_{i} \hat{n}_{i\downarrow} \hat{n}_{i\uparrow}
+ {1\over{2}} \sum_{i\neq j}V_{ij} \hat{\rho}_{i} \hat{\rho}_{j}~,\label{eq:Htqd}
\end{eqnarray}
where the operators $d_{i\sigma}$ ($d_{i\sigma}^\dag$) annihilate (create)
an electron with spin $\sigma=\pm 1/2$ on orbital $i$ ($i=1,2,3$).
$\hat{n}_{i\sigma} = d_{i\sigma}^\dag d_{i\sigma}$ and $\hat{\rho}_{i} = \hat{n}_{i\downarrow} + \hat{n}_{i\uparrow}$
are, respectively, the spin and charge density on orbital level $i$.
Each dot is represented by a single orbital with energy $\vep_{i\sigma}= \vep_i + g^* \mu_B B \sigma + \vep_0$,
with $g^*$ being the effective Land\'e $g$-factor and $\mu_B$ being the Bohr magneton.
$\vep_0$ is an overall energy shift with respect to the Fermi level of the leads
which can be changed by applying proper voltages to external gates.
Notice that $\vep_0$ changes the energy differences between states with different number of electrons,
but not the energy differences between the states with the same number of electrons.
$U_i$ is the on-site Coulomb repulsion of orbital $i$ and
$V_{ij}$ is the direct Coulomb interaction between two electrons in orbitals $i$ and $j$.
The effects of the perpendicular magnetic field on the hopping matrix
elements are accounted for by the Peierls phase factors\cite{peierls_zphys1933,luttinger_pr1951}
$t_{ij}(B)=t_{ij}e^{2\pi i \phi_{ij}}$.
For the three quantum dots located in the corners of an equilateral
triangle  we have $\phi_{12}=\phi_{23}=\phi_{31}=-\phi/3$ and
$\phi_{ji}=-\phi_{ij}$, where $\phi_B=BA/\phi_0$ is the number of magnetic
flux quanta threading the system, with $A$ being the area of the triangle
and $\phi_0=hc/e$ being the magnetic flux quantum.
The energy spectrum of the isolated TQD in the presence of the magnetic field
is obtained by the configuration-interaction method.\cite{delgado_shim_prb2007}

The TQD system ($D$) is connected to leads ($r$=$L,R$) on both sides
with left lead connected to dot 1 and right lead to dot 3 (Fig.~\ref{fig:system}).
We will use non-interacting one-dimensional chains as our model for leads
(see Appendix \ref{append:1Dchain} for explicit expressions
for the eigenstates and eigenvalues of the leads),
which are connected to the TQD by
\begin{eqnarray}\label{eq:HrD}
\widehat{H}_{rD}
&=& \sum_{\sigma} \sum_k \left( \tilde{t}_{rD}(k) c^{\dag}_{k\sigma} d_{i_0\sigma} + h.c. \right)~,
\end{eqnarray}
where $k$ is the eigenstates of the lead $r$, $\tilde{t}_{rD}(k)$ is the coupling strength
between the lead state $k$ and dot $i_0=1$ ($i_0=3$) for $r=L$ ($r=R$), given in Eq.~\eqref{eq:app.t_tilde}.
We will assume that tunnel coupling $t_{rD}$ between the lead site and the QD
is very small that we can use sequential tunneling picture described in the following.

\subsection{Transport theory and master equation}

We obtain the many-body eigenstates of the TQD molecule
by solving Eq.~\eqref{eq:Htqd} exactly for $N$=0 to 6.
The system of isolated TQD molecule and two leads is our unperturbed system,
and Eq.~\eqref{eq:HrD} is used as the perturbation
causing the transition between different states.
The tunneling between the leads and the quantum system is assumed to be small
so that we can consider only sequential tunneling transport.

The net current of electrons with spin $\sigma$ from the lead $r$ to TQD is
\beqa
I^{\sigma}_{r \rightarrow D}
&=& (-e) \left( \Omega_{r\rightarrow D}^{\sigma} - \Omega_{D\rightarrow r}^{\sigma} \right)~,
\eeqa
where $\Omega_{r\rightarrow D}^{\sigma}$ is the rate of
an electron with spin $\sigma$ moving from lead $r$ to the TQD, which is given by
\beqa
\Omega_{r\rightarrow D}^{\sigma}
&=& \sum_{N=0}^{5} \sum_{\alpha_N} \sum_{\alpha_{N+1}}
    f_r\left(E_{\alpha_{N+1}}^D-E_{\alpha_{N}}^D\right) P_{\alpha_N}
    \Gamma_r^{\sigma}\left(\alpha_N,\alpha_{N+1}\right)~,
\eeqa
and $\Omega_{D\rightarrow r}^{\sigma}$ is the rate of
an electron with spin $\sigma$ moving from TQD to lead $r$ which is given by
\beqa
\Omega_{D\rightarrow r}^{\sigma}
&=& \sum_{N=0}^{5} \sum_{\alpha_N} \sum_{\alpha_{N+1}}
    \left[1-f_r\left(E_{\alpha_{N+1}}^D-E_{\alpha_{N}}^D\right)\right] P_{\alpha_N+1}
    \Gamma_r^{\sigma}\left(\alpha_N,\alpha_{N+1}\right)~.
\eeqa
Here $f_r(\vep)=1/\{\exp[ (\vep-\mu_r)/k_BT ] +1 \}$ is
the Fermi-Dirac distribution function with respect to
the chemical potential $\mu_r$ of lead $r$ and the temperature $T$,
and $P_{\alpha_N}$ is the probability that the TQD
is in $N$-electron state $\alpha_N$ with energy $E^D_{\alpha_N}$.
$\Gamma^{\sigma}_r\left(\alpha_N,\alpha_{N+1}\right)$
is the transition rate of the TQD state from $\alpha_N$ to $\alpha_{N+1}$
by adding an electron with spin $\sigma$ from the lead $r$ to the TQD orbital $i_0$
which, by Fermi's Golden Rule, is
\beqa
\Gamma_r^{\sigma}\left(\alpha_N,\alpha_{N+1}\right)
&=& \frac{2\pi}{\hbar}
    \left| \langle \alpha_{N+1} | d^{\dag}_{i_0\sigma} | \alpha_N \rangle
    \right|^2
    \sum_k | \tilde{t}_{rD}(k) |^2
    \delta\left(  E_{\alpha_{N+1}}^D -  E_{\alpha_{N}}^D - \varepsilon_{k}^r \right)~.
\eeqa
The transition rate from $\alpha_{N+1}$ to $\alpha_N$ is also given
by the same $\Gamma_r^{\sigma}\left(\alpha_N,\alpha_{N+1}\right)$
and it describes the coupling between $\alpha_N$ and $\alpha_{N+1}$
by the connection to the lead $r$.
Individual transitions are assumed to be independent
and correlation between different many body states is
neglected in this formalism.\cite{density_matrix_approach}
The current of electrons with spin $\sigma$ from the lead $r$ to TQD can be expressed as
\beqa\label{eq:I_spectral}
I^{\sigma}_{r \rightarrow D}
&=& \frac{-e}{\hbar} \sum_{N=0}^5\sum_k \left| \tilde{t}_{rD}(k) \right|^2 \nonumber\\
&&\qquad\qquad\times\
    \Big\{ f_r\left(\vep^r_k\right) A_e\left(N;i_0,\sigma;\vep^r_k\right)
           -\left[1-f_r\left(\vep^r_k\right)\right] A_h\left(N+1;i_0,\sigma;\vep^r_k\right)
    \Big\}~,
\eeqa
where
\beqa
A_e\left(N;i,\sigma;\vep\right)
&=& 2\pi\sum_{\alpha_N}\sum_{\alpha_{N+1}} P_{\alpha_N}
    \left| \langle \alpha_{N+1} | d^{\dag}_{i\sigma} | \alpha_N \rangle
    \right|^2
    \delta\left(  E_{\alpha_{N+1}}^D -  E_{\alpha_{N}}^D - \vep \right)~, \label{eq:spectral_e}\\
A_h\left(N+1;i,\sigma;\vep\right)
&=& 2\pi\sum_{\alpha_N}\sum_{\alpha_{N+1}} P_{\alpha_{N+1}}
    \left| \langle \alpha_N | d_{i\sigma} | \alpha_{N+1} \rangle
    \right|^2
    \delta\left(  E_{\alpha_{N+1}}^D -  E_{\alpha_{N}}^D - \vep \right)~. \label{eq:spectral_h}
\eeqa
We refer to $A_e$ and $A_h$, respectively, as the electron and hole
spectral function of the TQD molecule.
These spectral functions contain information on the intrinsic properties of the TQD.
In transport, what is measured is essentially these spectral functions
weighted with the distribution functions of the leads $f_r$
and the lead-dot coupling $\tilde{t}_{rD}(k)$.
Note that the probabilities of each state are determined by the master equation described below,
and therefore the spectral functions depend on the bias.
The first term in Eq.~\eqref{eq:I_spectral} describes
an electron moving from the lead $r$ to the TQD system
increasing number of electrons in the TQD from $N$ to $N+1$.
The second term describes the removal of an electron
from the TQD with $N+1$ electrons back to the same lead $r$.

At equilibrium, the probabilities $P_{\alpha_N}$'s are given by
\beqa
P_{\alpha_N}^{\rm eq}
 &=& \frac{ \exp\left( -\frac{ E_{\alpha_N}^D - \mu_0 N }{k_B T} \right)}{Z}~,\label{eq:P_eq}
\eeqa
where $Z$ is the grand partition function of the TQD molecule
and $\mu_0$ is the chemical potential at equilibrium.
When a finite bias is applied between the left and right leads,
the time evolution of the probabilities $P_{\alpha_N}$ is given by the following set of master equations.
\begin{eqnarray}\label{eq:Master}
\frac{dP_{\alpha_{N}}}{dt}
&=&\; \sum_{\alpha_{N+1}} \sum_{r=L,R} P_{\alpha_{N+1}}\left[ 1-f_r\left(E_{\alpha_{N+1}}^D -  E_{\alpha_{N}}^D\right) \right]
                          \Gamma_r(\alpha_{N},\alpha_{N+1}) \nonumber\\
&&   -\sum_{\alpha_{N+1}} \sum_{r=L,R} P_{\alpha_{N}} f_r\left(E_{\alpha_{N+1}}^D -  E_{\alpha_{N}}^D\right)
                          \Gamma_{r}(\alpha_N,\alpha_{N+1}) \nonumber\\
&&   +\sum_{\alpha_{N-1}} \sum_{r=L,R} P_{\alpha_{N-1}} f_r\left(E_{\alpha_{N}}^D -  E_{\alpha_{N-1}}^D\right)
                          \Gamma_{r}(\alpha_{N-1},\alpha_{N}) \nonumber\\
&&   -\sum_{\alpha_{N-1}} \sum_{r=L,R} P_{\alpha_{N}}\left[ 1-f_r\left(E_{\alpha_{N}}^D -  E_{\alpha_{N-1}}^D\right) \right]
                          \Gamma_r(\alpha_{N-1},\alpha_{N})~,
\end{eqnarray}
where $\Gamma_r = \sum_{\sigma}\Gamma_r^{\sigma}$.
The first and third terms are the contribution from the transitions
$\alpha_{N\pm 1} \rightarrow \alpha_N$,
and the second and fouth terms are the contribution from the transitions
$\alpha_N \rightarrow \alpha_{N\pm 1}$.
For $N=0$($N=6$), the summations over $\alpha_{N-1}$($\alpha_{N+1}$) is absent in the master equation.
If we define a vector $\mathbf{P}$ whose elements are $P_{\alpha_N}$ of all possible $N$ and $\alpha_N$,
the master equation can be cast into a matrix form:
\begin{equation}
\frac{d\mathbf{P}}{dt} = \mathbf{M}_P\cdot\mathbf{P}~,\label{eq:Master_matrix}
\end{equation}
where the matrix $\mathbf{M}_P$ is defined by Eq.~\eqref{eq:Master}.
The initial probabilities at $t=0$ are given by equilibrium values
$P_{\alpha_N}(0) = P_{\alpha_N}^{\rm eq}$.
The normalization condition that the sum of all the probabilities is unity,
\beqa
&& \sum_N \sum_{\alpha_N} P_{\alpha_N} = 1~,\label{eq:normalization_P}
\eeqa
is conserved by the master equation.

At steady state, all the time derivatives are zero
and the master equation Eq.~\eqref{eq:Master_matrix} becomes a homogeneous system of linear equations,
\beqa
\mathbf{M}_P\cdot\mathbf{P} &=& 0~.\label{eq:Master_steady}
\eeqa
We can solve this equation combined with the normalization condition Eq.~\eqref{eq:normalization_P}.
This equation has nontrivial solutions and
sometimes the solution can not be uniquely determined with only the normalization condition.
To uniquely determine the steady state probabilities, we need to make use of the initial conditions.
Using the singular value decomposition of the matrix $\mathbf{M}_P$,
we can uniquely determine the steady state probabilities
(see Appendix~\ref{append:steady_sol} for details).

\section{Conditions for transport through a quantum many-body system}\label{sec:Transport}

Before we move on to the numerical results, we discuss the conditions
required for transport through a general quantum system connected by two leads at both ends.
We will call the general quantum system ``dot'' in this section.
The transport in general involves multiple many-body quantum states
and can not be reduced to a single electron description.
Transport channels for this many-body, many-channel transport in the sequential tunneling regime
will be defined and general conditions for transport will be given.
We will assume zero temperature to simplify the discussion.

Without source-drain bias ($V_{\rm sd}$=0), the dot will be in equilibrium
with the leads with chemical potential $\mu_0$.
The number of electrons in the dot at equilibrium will be determined by
\beq
\mu_D(N-1) < \mu_0 < \mu_D(N)~,
\eeq
where $\mu_D$ is the chemical potential of the dot defined as
\beq
\mu_D(N) \equiv E^D_{\alpha_{N+1}^{GS}}-E^D_{\alpha_{N}^{GS}}~,
\eeq
for the $N$-electron ground state $\alpha_{N}^{GS}$
and ($N+1$)-electron ground state $\alpha_{N+1}^{GS}$.
The dot will be in the ground state of $N$-electron system, $\alpha_{N}^{GS}$.
Once we apply a bias, transition from the ground state $\alpha_{N}^{GS}$ to other states will occur.
By succession of such transitions, electrons can move from one lead to the other lead.
During this process many of the dot states will be accessed
and on the average we can assign a probability
that the dot will be in each many-body state.
Probabilities will change in time initially, but reach stationary values at steady state.

To be specific, let us assume that we apply a bias $V_{\rm sd}$ between the two leads
such that $\mu_L = \mu_0 + eV_{\rm sd}/2$ and $\mu_R = \mu_0 - eV_{\rm sd}/2$.
For a transition from an $N$-electron state $\alpha_N$
to an ($N+1$)-electron state $\beta_{N+1}$
by adding an electron in the dot from lead $r=L,R$ to occur,
the incoming electron must have the energy of $E^D_{\beta_{N+1}}-E^D_{\alpha_N}$.
Electrons in the lead $r$ have energies below the chemical potential $\mu_r$ and
therefore the condition $E^D_{\beta_{N+1}}-E^D_{\alpha_N} \leq \mu_r$ must be satisfied.
In addition, the states $\alpha_N$ and $\beta_{N+1}$
must be connected by adding an electron from the lead $r$,
{\it i.e.}, $\Gamma_r (\alpha_N, \beta_{N+1}) \neq 0$.
Similarly, a transition from an ($N+1$)-electron state
$\beta_{N+1}$ to an $N$-electron state $\alpha_N$
by moving an electron from the dot to the lead $r$ is allowed if
$E^D_{\beta_{N+1}}-E^D_{\alpha_N} \geq \mu_r $ and $\Gamma_r (\alpha_N, \beta_{N+1}) \neq 0$.
In this case, there must be an empty state
with energy $E^D_{\beta_{N+1}}-E^D_{\alpha_N}$ in the lead $r$.
We define a set of two successive transitions
\beq
\alpha_N \stackrel{L}{\rightarrow} \beta_{N+1} \stackrel{R}{\rightarrow} \alpha_{N}'
\quad \text{or}\quad
\beta_{N+1} \stackrel{R}{\rightarrow} \alpha_{N} \stackrel{L}{\rightarrow} \beta_{N+1}' \nonumber
\eeq
as a {\it transport channel} if ({\it i}) these transitions are allowed
and ({\it ii}) the initial state $\alpha_N$ or $\beta_{N+1}$ is either the ground state
or accessible from the ground state by successive allowed transitions.
The $L$ or $R$ on top of the arrows means the transition is allowed by connection to the left or right lead.
If two states are involved in a transport channel such as $\alpha_N \rightarrow \beta_{N+1} \rightarrow \alpha_N$,
we will simply call the pair $(\alpha_N,\beta_{N+1})$ a transport channel.
We will define any state that participates in one or more transport channels an {\it active state}.
In most cases, if there are one or more transport channels, the current flows through the system.
Exception is when there is a trap state.
A state $\gamma_N$ is defined to be a {\it trap state} if
$(i)$ $\gamma_N$ is accessible from the ground state through successive allowed transitions
and $(ii)$ the transition $\gamma_N \rightarrow \alpha_{N\pm 1}$
is not allowed for any active state $\alpha_{N\pm 1}$.
When we apply a bias to the system, the probability of the trap state monotonically increases in time
and in the steady state condition, the system will be completely in the trap state and current can no longer proceed.
The (1,1) triplet state in DQD spin-blockade system \cite{ono_austing_science2002}
with forward bias is an example of trap state.
To summarize, the general conditions for transport via sequential tunneling are
$(i)$ there is at least one transport channel and
$(ii)$ there is no trap state.

We can understand the expression for total current in Eq.~\eqref{eq:I_spectral}
in terms of transport channels, active states and trap states.
If there is no trap state, then each active state will have a finite probability.
The electron spectral function [Eq.~\eqref{eq:spectral_e}] is
sum over all transitions from an active state $\alpha_N$ to any state $\alpha_{N+1}$.
For the total current, this electron spectral function is weighted by the Fermi-Dirac function.
At zero temperature, this weighted spectral function is sum over
all transitions only between active states.
Similarly, the hole spectral function is sum over all transitions
from an active state $\alpha_{N+1}$ to any $N$-electron state.
It is weighted with $1-f_r$ and the weighted hole spectral function
is sum over all transitions between active states.
If there is a trap state $\gamma_N$, then only $P_{\gamma_N}$(=1) is nonzero
and the weighted spectral function is sum over transitions from $\gamma_N$ to other active states.
Since there is no allowed transition from $\gamma_N$
to any active state by definition of the trap state,
the weighted spectral function is zero and there is no current.

\section{Results}\label{sec:Results}

We use the effective Rydberg $Ry$ and the effective Bohr radius $a_B$
of the host semiconductor material as our units for energy and length, respectively.
For GaAs, $Ry=5.93$ meV and $a_B=9.79$ nm.
The three QDs are located at the vertices of a equilateral triangle
and the inter-dot distance is $6.25$.
The magnetic field is measured with the number of flux quanta $\phi_B$ through the triangle.
We use parameters $U_i=U=2.5$, $V_{ij}=V=0.5$, $t_{ij}=-t=-0.05$, for all $i$ and $j$.
The tunneling strengths $t_L$ and $t_R$ in the leads
are assumed to be large so that the density of states of the leads
are nonzero in a wide range of energy. We use $t_L=t_R=-4.0$ in the calculation.
The current is calculated in units of $I_0=e|t_{LD}|^2/\hbar |t_L|$,
and conductance in units of $G_0=e^2|t_{LD}|^2/\hbar |t_L|$.
We assume that $t_{LD}=t_{RD}$
and that they are small enough to justify the sequential tunneling picture.
In these effective units, the current and the conductance
do not depend on the dot-lead tunneling $t_{LD}$ and $t_{RD}$.
The effective $g$-factor is taken to be $-0.44$.
The chemical potentials of each leads are
$\mu_L = \mu_0 + eV_{\rm sd}/2$ and $\mu_R = \mu_0 - eV_{\rm sd}/2$.
We set the equilibrium chemical potential $\mu_0$ to be zero.
All the calculations are done at temperature $T$=0.001.

\subsection{Quadruple point with a trapped hole}

We consider a quadruple point where a hole is trapped in dot 2.
A hole in the TQD is defined as the absence of an electron
with respect to the fully occupied state with six electrons.
At this quadruple point, the four resonant charge configurations are
(2,1,2),(1,1,2),(2,0,2) and (2,1,1).
If we neglect the tunneling between the constituent dots of the TQD molecule,
the QP can be determined by equating the energies of the four charge configurations,
\beq\label{eq:EQP}
E(2,1,2)=E(1,1,2)=E(2,0,2)=E(2,1,1)\equiv E_{QP}~.
\eeq
With $\vep_1=\vep_3=\vep$ and $\vep_2=\vep+\Delta$, the energies of the four configurations are
\beqa
E(2,1,2) &=& 5\vep+\Delta+2U+8V~, \\
E(1,1,2) &=& E(2,1,1) = 4\vep+\Delta+U+5V~, \\
E(2,0,2)&=& 4\vep+2U+4V~,
\eeqa
and from Eq.~\eqref{eq:EQP} we obtain $\vep=-U-3V$ and $\Delta=U-V$.
At this classical QP the level energies of the QDs are
\beqa
\vep_1&=&\vep_3=-U-3V~, \\
\vep_2&=&-4V~,
\eeqa
and the total energy is
\beq
E_{QP}=-2U-8V~.
\eeq
When we have tunneling between the QDs,
the charge configurations in terms of the population of each individual dots
are not eigenstates of the coupled TQD molecule system.
The quadruple points are defined then as the point
where all the resonant charge configurations have the same probabilities,
which can be determined numerically.
The QP with a trapped hole where the four charge configurations
(2,1,2),(1,1,2),(2,0,2), and (2,1,1) have same probabilities
corresponds to the level energies
$\vep_1=\vep_3=-4.05$ and $\vep_2=-2.095$ with $\vep_0=0$.
Notice that these values are quite close to the values of the classical QP
because the tunneling $t$ is relatively small compared to the Coulomb interaction parameters $U$ and $V$.

In terms of hole occupation numbers,
the charge configurations (0,1,0)$_h$, (1,1,0)$_h$, (0,2,0)$_h$, and (0,1,1)$_h$
are degenerate where ($N_1,N_2,N_3$)$_h$=($2-N_1,2-N_2,2-N_3$).
The lower part of the energy spectrum of the TQD at this QP is shown in Fig.~\ref{fig:QP2_spectrum}(a).
The two thin black dotted lines are the energy levels of spin-up and spin-down states of the trapped hole.
There is no AB oscillations for the single-hole states because it is localized in dot 2.
For two-hole states, neglecting very high energy states
consisting mostly of configurations (2,0,0)$_h$, (0,0,2)$_h$ and (1,0,1)$_h$,
we have nine two-hole states (3 singlet and 6 triplet states).
Two-hole singlet states (red solid curves) show strong AB oscillations
since they consist of three configurations (1,1,0)$_h$, (0,2,0)$_h$, and (0,1,1)$_h$.
That is, there is one hole trapped in dot 2 and an extra hole moves around the three dots.
Two-hole triplet states mainly consist of only two configurations (1,1,0)$_h$ and (0,1,1)$_h$
because (0,2,0)$_h$ configuration can be only singlet due to the exclusion principle.
Therefore, triplet states (blue dashed curves) show only small AB oscillations
which can be ascribed to higher energy triplet configuration (1,0,1)$_h$.
This suppression of AB oscillation in triplet states also
occurs at the QP with a trapped electron in dot 2.\cite{delgado_shim_prl2008}
The energy spectrum of the TQD with a trapped hole has two main differences
compared to the TQD with a trapped electron. First, at zero magnetic field,
two-hole spin-singlet and triplet states are almost degenerate.
Second, the phase of the AB oscillation of the singlet state is shifted by $\pi$.
Without Zeeman splitting, these differences
would lead to $\pi$ phase-shifted AB oscillations with less amplitudes in current
because the triplet transport channel will have finite contribution with very small AB oscillation.
With Zeeman splitting, the single-hole and two-hole ground states
are on resonance only at zero magnetic field.
We can tune the relative energy difference between two-hole ({\it i.e.}, four-electron) and
single-hole ({\it i.e.}, five-electron) states by changing the overall energy shift $\vep_0$.
The two-hole states obtain $4\vep_0$ while single-hole states obtain $5\vep_0$.
Therefore, single-hole states gain additional energy $\vep_0$ with respect to two-hole states.
It is easier to understand how the transport channels contribute to the total current
if the one-hole and two-hole states are energetically separated.
Figure \ref{fig:QP2_spectrum}(b) is the energy spectrum with $\vep_0$=-0.15,
where the single-hole states are well below the two-hole states.
Interesting phenomena such as negative differential conductance
and spin-selective AB oscillations occur in this system.
We will present results for this system at various biases and magnetic fields.

\subsection{Spectral functions and transport channels}

\begin{figure}
  \includegraphics[width=0.8\linewidth]{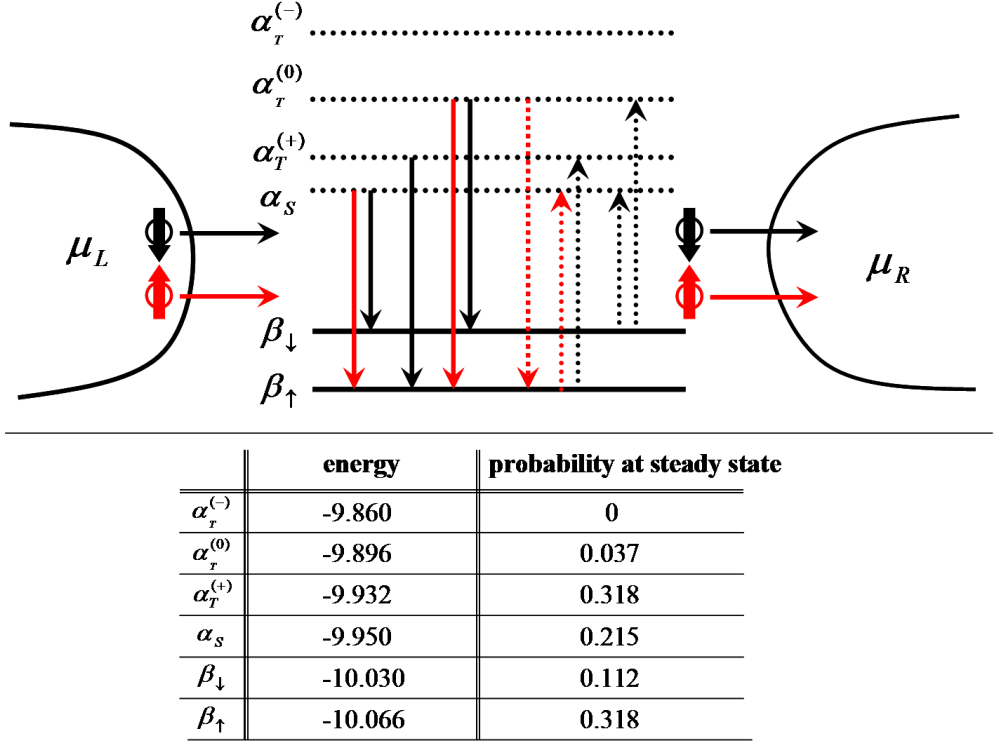}\\
  \caption{(Color online) Schematic picture describing the transitions at $\phi_B$=3.3 and $eV_{\rm sd}$=0.3.
           $\beta$'s are single-hole levels with spin up and down, and $\alpha$'s are two-hole levels.
           $\alpha_S$ is spin singlet and $\alpha_T^{(\pm,0)}$ are spin triplet.
           Solid (Dotted) arrows are transitions allowed by connection to the left (right) lead.
           Transitions represented by downward arrows are from a two-hole state to
           a single-hole state by adding an electron from the lead,
           and upward transitions are from a single-hole state to a two-hole state
           and emit an electron to the lead.
           The black (red) color represents that the spin
           of the electron transporting in the transition is down (up).
           Energies and the probabilities of each level are also given for reference.}
  \label{fig:transitions}
\end{figure}

To explain how the current is related to the spectral functions and the transport channels,
we choose a specific case with $\phi_B$=3.3 and $eV_{\rm sd}$=0.3.
The system is schematically given in Fig.~\ref{fig:transitions}.
The single-hole states with spin up ($\beta_{\uparrow}$)
and down ($\beta_{\downarrow}$) are the lowest levels.
For two-hole states, $\alpha_S$ is spin singlet
and $\alpha_T^{(\pm,0)}$ is spin triplet with total $S_z$=$\pm 1$, 0 respectively.
Solid black (red) arrows represent transitions from two-hole states to single-hole states
by a spin-down (spin-up) electron moving from the left lead to the TQD system.
These transitions satisfy
\beq
E^D_{\beta}-E^D_{\alpha} \le \mu_L = \mu_0+\frac{eV_{\rm sd}}{2} = 0.15 ~,
\eeq
and $\Gamma_L(\alpha,\beta)\neq 0$.
$\alpha_T^{(-)}$ and other higher energy levels are
outside the transport window and are not accessed during transport.
Notice that the transition $\alpha_T^{(+)}\rightarrow\beta_{\downarrow}$ is energetically possible
but not allowed by spin-blockade since adding a single electron cannot change the total spin $S_z$ from +1 to -1/2.
The transition from $\alpha_T^{(0)}$ to $\beta_{\uparrow}$ can
also occur by adding an electron from the right lead since the incoming electron must have energy -0.17,
which is below the chemical potential of the right lead $\mu_R$=-0.15.
This is the only transition that transports an electron from the right lead to the TQD,
and is represented as a dotted red arrow pointing downward.
Dotted black (red) arrows pointing upward represent transitions from single-hole states to two-hole states
by a spin-down (spin-up) electron moving from the TQD system to the right lead.
These transitions satisfy
\beq
E^D_{\beta}-E^D_{\alpha} \ge \mu_R = \mu_0-\frac{eV_{\rm sd}}{2}=-0.15~,
\eeq
and $\Gamma_R(\alpha,\beta)\neq 0$.
There are four pairwise transport channels $(\alpha_S,\beta_{\uparrow})$,
$(\alpha_S,\beta_{\downarrow})$, $(\alpha_T^{(+)},\beta_{\uparrow})$, $(\alpha_T^{(0)},\beta_{\downarrow})$
as well as other transport channels such as
$\beta_{\uparrow}\rightarrow\alpha_S\rightarrow\beta_{\downarrow}$, etc.
There is no trap state in this case.

\begin{figure}
  \includegraphics[width=0.6\linewidth]{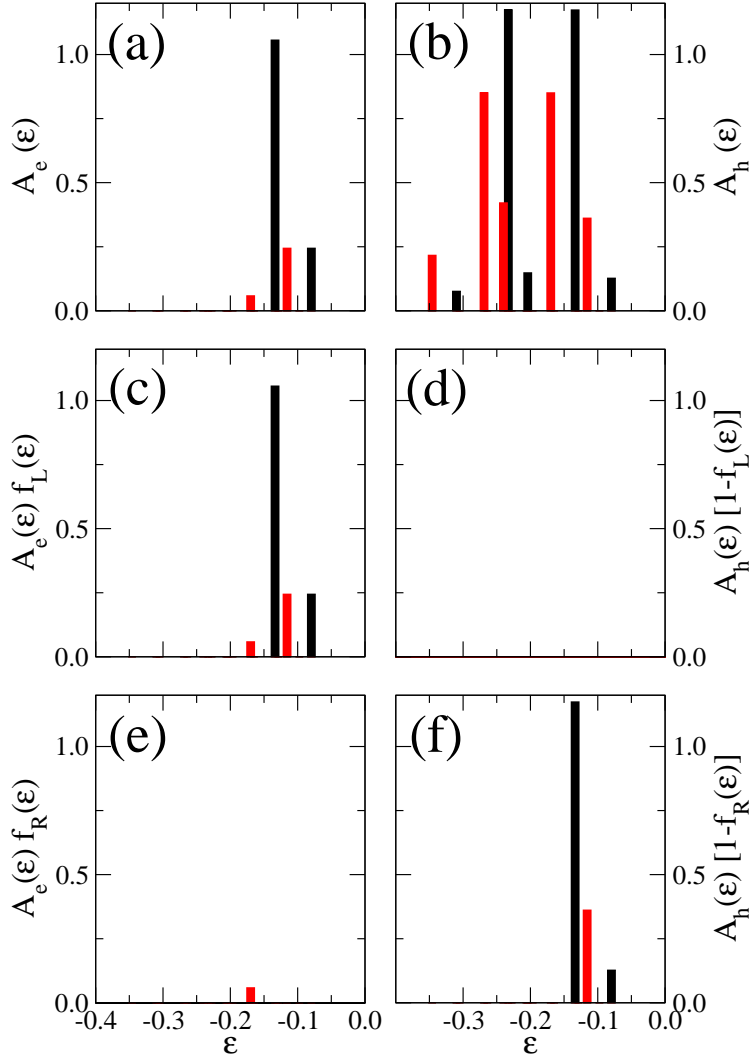}\\
  \caption{(Color online) Spectral functions at $\phi_B$=3.3 and $eV_{\rm sd}$=0.3.
            Black (red) columns are for $\sigma$=$\downarrow$ ($\sigma$=$\uparrow$).
           (a) Electron spectral function, $A_e(N\text{=4};i,\sigma;\vep)$.
           (b) Hole spectral function, $A_h(N\text{=5};i,\sigma;\vep)$.
           $i$=1 or 3 gives the same spectral functions since we have symmetric system
           and assumed $t_{LD}$=$t_{RD}$.
           (c) and (e) are electron spectral functions weighted by the Fermi function $f_L$ and $f_R$
           respectively, and represent allowed transitions from a two-hole state to a single-hole state
           by an electron moving from the lead to the TQD system.
           (d) and (f) are hole spectral functions weighted by 1-$f_L$ and 1-$f_R$ respectively,
           and represent allowed transitions from a single-hole state to a two-hole state
           by an electron moving from the TQD system to the lead.
           No transition moves an electron from the TQD to the left lead
           and therefore there is no visible peaks in (d).
           These weighted spectral functions show all the transitions during the transport
           depicted by arrows in Fig.~\ref{fig:transitions}. }
  \label{fig:spectral_V0.3}
\end{figure}

The contribution of each transition to the current can be understood by spectral functions.
Figure \ref{fig:spectral_V0.3} shows the spectral functions for this system.
Due to the symmetry of the system and the assumption $t_{LD}$=$t_{RD}$,
spectral functions for $i$=1 and $i=3$ are the same.
Spectral functions are sum of delta-functions and
the coefficients of each delta-function are plotted here.
(a) is the electron spectral function and (b) is the hole spectral function.
Electron (Hole) spectral function shows transitions
from any active two-hole (single-hole) state to any single-hole (two-hole) state, not necessarily only to active states.
But not all these transitions actually occur during the transport.
Only those transitions within the transport window occur and contribute to the current.
For the current from the left lead to the TQD,
the summation is over the electron spectral function
weighted by the Fermi function of the left lead.
Black (Red) columns in (c) show the weighted electron spectral function
for adding a spin-down (spin-up) electron,
which correspond to the solid black (red) arrows in Fig.~\ref{fig:transitions}.
Transitions $\alpha_T^{(+)}\rightarrow \beta_{\uparrow}$ and
$\alpha_T^{(0)}\rightarrow \beta_{\downarrow}$ are induced by spin-down electrons with same energy
and the highest black column represents sum of these two.
This weighted electron spectral function is the same
with the bare spectral function in the energy range shown here
because all the transitions from active two-hole states
to any of the $\beta_{\uparrow}$ and $\beta_{\downarrow}$ are in the transport window.
For the current from the TQD to the left lead,
the hole spectral function is weighted by 1-$f_L$ [(d)].
This weighted hole spectral function is zero
since no transition from a single-hole state to a two-hole state
emits an electron with energy larger than the chemical potential of the left lead
(there is no solid arrows pointing upward in Fig.~\ref{fig:transitions}).
For the current from the TQD to the right lead,
we need to sum over the hole spectral function weighted by $1-f_R$.
Black (Red) columns in (f) correspond to the dotted black (red) arrows
pointing upward in Fig.~\ref{fig:transitions}.
For the current from the right lead to the TQD,
the electron spectral function is weighted by $f_R$ [(e)],
which has only one small peak corresponding to the transition
$\alpha_T^{(0)}$ to $\beta_{\uparrow}$ represented
by the dotted red arrow pointing downward in Fig.~\ref{fig:transitions}.
These weighted spectral functions show the transitions between active states
({\it i.e.}, transitions shown in Fig.~\ref{fig:transitions})
and only these transitions form transport channels and contribute
to the current.
The net current from the left lead to the TQD $I_{L\rightarrow D}$ can be obtained
by summing the peaks in (c) and
the net current from the TQD to the right lead $I_{D\rightarrow R}$
is the sum of the peaks in (f) minus the peak in (e).
The net currents $I_{L\rightarrow D}$ and $I_{D\rightarrow R}$
are the same, satisfying the steady state condition.

\subsection{Negative differential conductance}

When we increase the bias, the transport window expands and
more transport channels are involved in the transport.
Whenever a new transport channel is introduced, the probabilities
of each states must be redistributed
and the spectral functions and the current change accordingly.
In the range of biases where no new transport channel is introduced
the current remains flat and therefore nonzero differential conductance
signifies introduction of new transport channel.
Figure~\ref{fig:QP2_DV_phi_IGP} shows how the currents[(a) and (d)],
differential conductances[(b) and (e)],
and the probabilities[(c) and (f)] of each states change
for the same system as in Fig.~\ref{fig:QP2_spectrum}(b)
as we increase the bias at two different magnetic fields $\phi_B$=0 and 3.3.
At zero magnetic field [(a)$\sim$(c)], the single-hole state is doubly spin-degenerate
and the current changes when a two-hole state enters the transport window as we increase the bias.
At each current plateaus the probabilities of each active states are the same,
which results from the assumption that $t_{LD}$=$t_{RD}$.
The differential conductance peaks at the energy differences between
two-hole states and the single-hole ground state.
At finite magnetic field $\phi_B$=3.3 [(d)$\sim$(f)],
the spin-degeneracy of the single-hole states is lifted
and we have multiple $N$- and ($N+1$)-particle states.
The current changes whenever additional transport channel is introduced by increasing bias,
which is more complicated than at zero magnetic field
since the two single-hole levels have different energies.
The probabilities of each active states at current plateaus are not the same in this case.
In most cases the additional transport channel results in the increase in current,
but sometimes it leads to the decrease in the current
as is the case for the decrease in current
at $eV_{\rm sd} \simeq 0.7$ for $\phi_B=0$ [Fig.~\ref{fig:QP2_DV_phi_IGP}(a)],
and at $eV_{\rm sd} \simeq$ 0.62 and 0.69 for $\phi_B=3.3$ [Fig.~\ref{fig:QP2_DV_phi_IGP}(d)].
This negative differential conductance can be explained as a result of the interplay
between the different coupling strength $\Gamma_r(\alpha,\beta)$ for singlet and triplet states
and the probability redistribution of many-electron states with increased bias.
The single-hole states are more strongly coupled to the triplet two-hole states
than to the singlet two-hole states ($\Gamma_r$ is larger for the triplet)
because the triplet states consist of two configurations [(1,1,0)$_h$ and (0,1,1)$_h$]
while the singlet states consist of three configurations [(1,1,0)$_h$, (0,2,0)$_h$, and (0,1,1)$_h$],
and the configuration (0,2,0)$_h$ is not connected to the single hole configuration (0,1,0)$_h$
by adding or subtracting a hole from the leads.
If increasing the bias makes the singlet state above the triplet states
become an active state, we have more transport channels.
But the probabilities of all the active states will be redistributed
and the triplet active states will have less probabilities.
If the current decrease due to the reduction of the triplet probabilities
is larger than the current increase due to
the introduction of new transport channels involving the singlet state,
the total current decreases as we increase the bias.

\begin{figure}
  \includegraphics[width=0.8\linewidth]{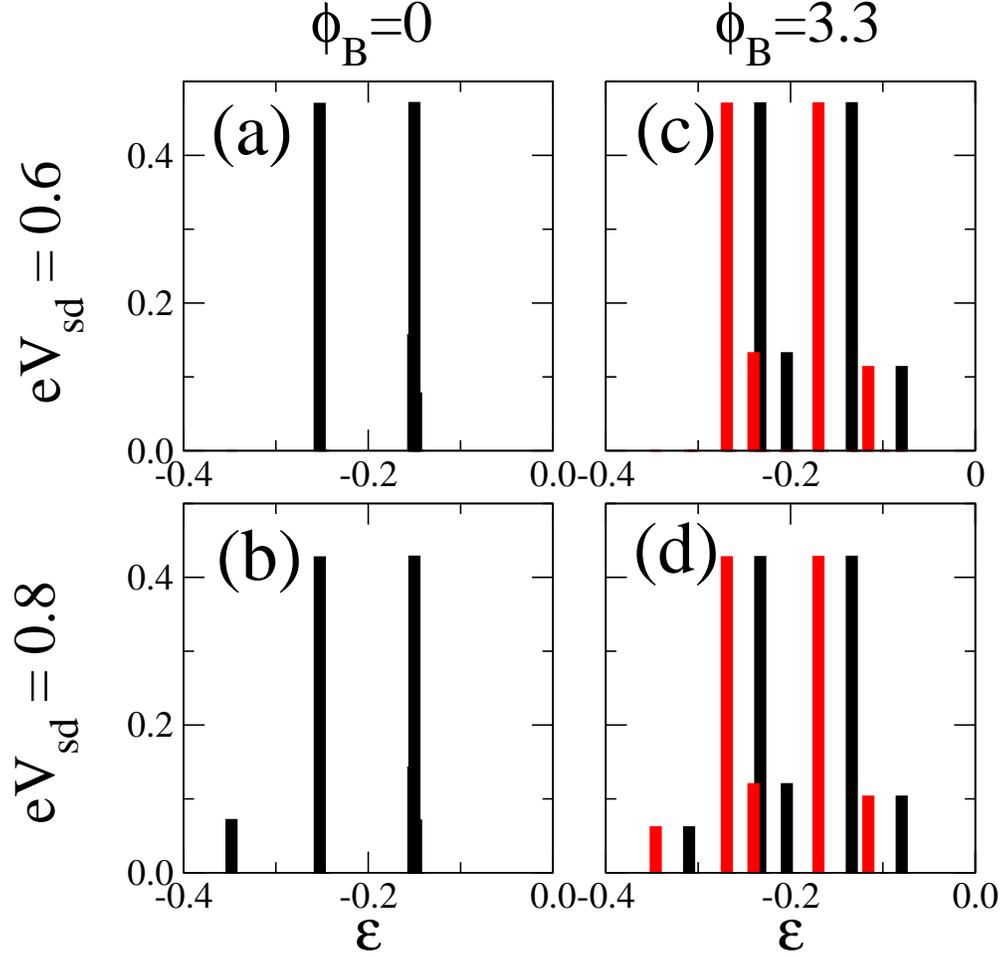}\\
  \caption{(Color online) Electron spectral functions weighted by $f_L$
            before [(a) and (c)] and after [(b) and (d)] current decrease.
            At zero field [(a) and (b)],
            spin-down and spin-up components are the same due to the spin degeneracy.
            At finite field [(c) and (d)], more peaks appear due to the lifting of the spin degeneracy.
            The introduction of the transport channels involving the high-energy singlet state
            as the bias increases leads to redistribution of the probabilities
            and reduction of the current through the triplet transport channels which are dominant channels.
            The total current decreases since the decrease through the triplet transport channels is bigger than
            the increase through the new singlet transport channels.}
  \label{fig:spectral_NDR}
\end{figure}

Figure \ref{fig:spectral_NDR} shows the electron spectral functions weighted by $f_L$
before [(a) and (c)] and after [(b) and (d)] the current decrease.
The electron spectral functions weighted by 1-$f_L$ are all zero for this case,
and the current is simply proportional to the sum of the heights of all columns in the figure.
At zero magnetic field [(a) and (b)] we show only spin-down component because
spin-up component is the same due to the spin degeneracy.
Before the current decrease, at $eV_{\rm sd}$=0.6 [(a)],
the two high columns correspond to the transitions from spin-triplet states to single-hole states
and the two small columns which almost overlap with one of high column
are from the transitions from spin-singlet states.
The three columns are very close together due to the singlet-triplet degeneracy of the two-hole system.
After the current decrease, at $eV_{\rm sd}$=0.8 [(b)],
there is additional peak at around $\vep\sim -0.35$, which correspond to the new allowed transitions
involving the higher singlet state.
Notice that the decrease of the two peaks of triplet transitions
is bigger than the new singlet peak,
which leads to the negative differential conductance.
At finite magnetic field [(c) and (d)],
spin degeneracy is lifted and there are more peaks in the weighted spectral function.
The higher peaks correspond to transitions from triplet states
and lower peaks correspond to transitions from singlet states.
At $eV_{\rm sd}$=0.8 [(d)], the two new peaks correspond to
the transitions from the third singlet state to the two single-hole states.
Once again, the peaks of triplet transport channels decrease
and the total current decreases.

\subsection{Spin-selective Aharonov-Bohm oscillations}

\begin{figure}
  \includegraphics[width=0.6\linewidth]{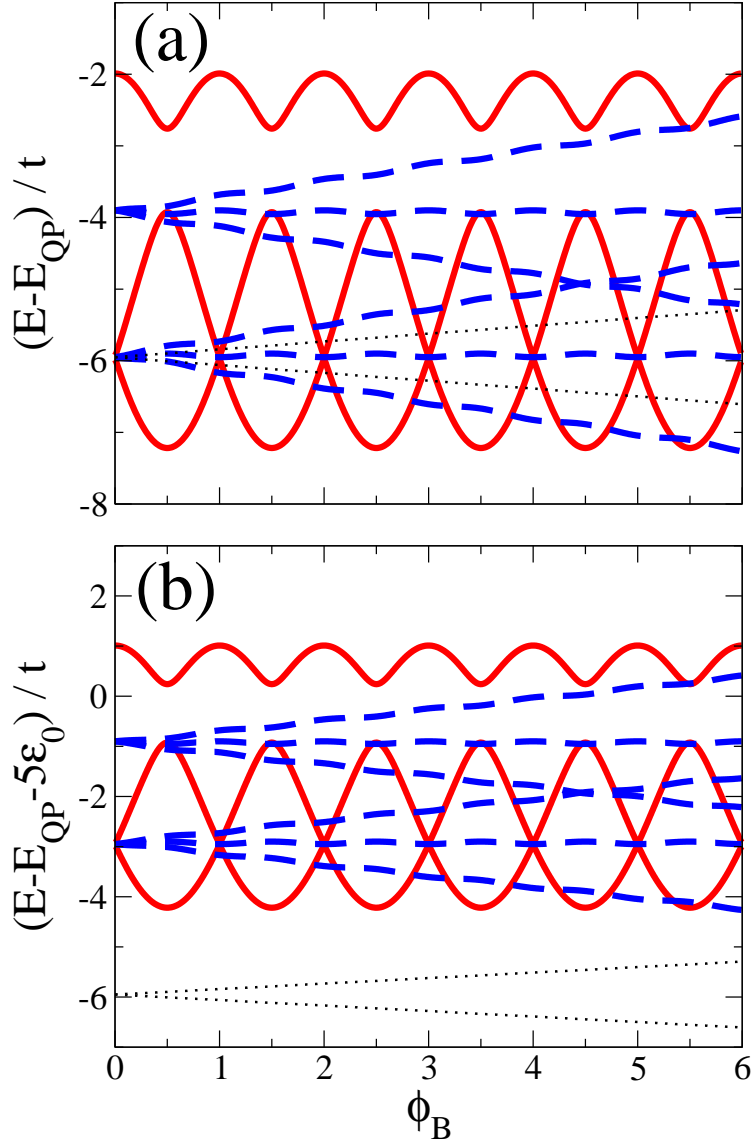}\\
  \caption{(Color online) (a) Energy spectrum at the QP where a hole is trapped in dot 2.
            Four charge configurations (0,1,0)$_h$, (1,1,0)$_h$, (0,2,0)$_h$, and (0,1,1)$_h$
            in terms of the hole occupation numbers have the same probability at zero magnetic field.
            Black dotted curves are for single-hole states with spin up and down.
            Red solid (Blue dashed) curves are spin-singlet (spin-triplet) states for two-hole systems.
            (b) Energy spectrum with the overall shift $\vep_0$=-0.15.
            By changing $\vep_0$, the single-hole states move upward for positive $\vep_0$
            and downward for negative $\vep_0$ with respect to the two-hole states.}
  \label{fig:QP2_spectrum}
\end{figure}

\begin{figure}
  \includegraphics[width=0.6\linewidth]{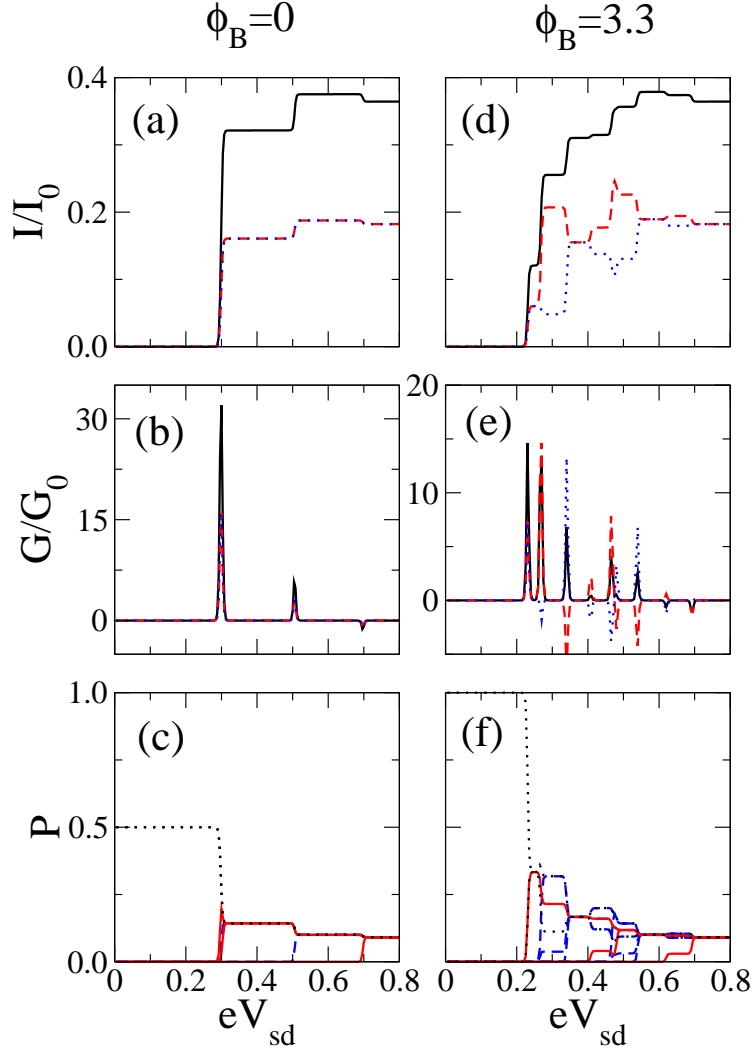}\\
  \caption{(Color online) Currents, differential conductance, and probabilities as functions of the bias
            at two different magnetic fields $\phi_B$=0 [(a)$\sim$(c)] and $\phi_B$=3.3 [(d)$\sim$(f)],
            for the same system as in Fig.~\ref{fig:QP2_spectrum}(b).
            For the currents and conductances, black solid curves are the total currents or conductances and
            red dashed (blue dotted) curves are for spin down (up) components.
            For the probabilities, black dotted curves are for single-hole states and
            red solid (blue dashed) curves are for spin-singlet (spin-triplet) states.}
  \label{fig:QP2_DV_phi_IGP}
\end{figure}

\begin{figure}
  \includegraphics[width=0.6\linewidth]{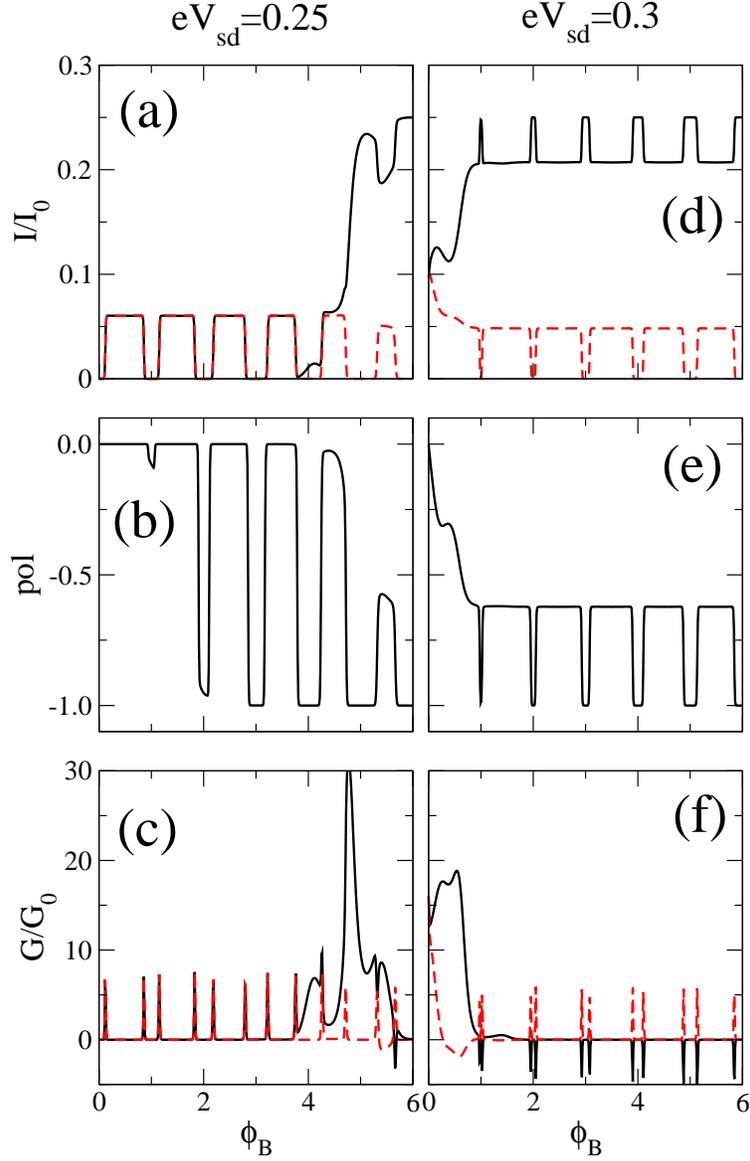}\\
  \caption{(Color online) Spin-selective AB oscillations at two different values of bias,
           $eV_{\rm sd}$=0.25 [(a)$\sim$(c)] and 0.3 [(d)$\sim$(f)].
           Top panels show the spin-down (black solid curves) and spin-up (red dashed curves) currents.
           Bottom panels show the differential conductance of each spin species.
           Middle panels show the spin polarization of the current, defined as
           $pol=(I_{\uparrow}-I_{\downarrow})/(I_{\uparrow}+I_{\downarrow})$.}
  \label{fig:QP2_spinAB_finiteV}
\end{figure}

The different oscillatory behavior of singlet and triplet states in the magnetic field
can lead to spin-dependent transport phenomena.
Left panels of Fig.~\ref{fig:QP2_spinAB_finiteV} [(a)$\sim$(c)] show
the current, current polarization, and differential conductance as a function of the magnetic field
at $eV_{\rm sd}$=0.25 for the same system as in Fig.~\ref{fig:QP2_spectrum}(b).
At this bias, the lowest singlet two-hole state ($\alpha_S$)
can form transport channels with both the spin-up ($\beta_{\uparrow}$) and
spin-down ($\beta_{\downarrow}$) single-hole state at lower magnetic fields.
$\beta_{\uparrow}$ and $\beta_{\downarrow}$ have the same probabilities in this regime and
the spin-up and spin-down components of current are the same and the spin-polarization of the current is zero.
The AB oscillation makes the singlet state $\alpha_S$ oscillate in and out of the transport window,
which leads to the oscillations in current and conductance.
At higher magnetic fields, the spin-triplet state with $S_z$=+1 ($\alpha_T^{(+)}$)
comes into the transport window with respect to $\beta_{\uparrow}$ and $\beta_{\downarrow}$.
Since $\beta_{\downarrow}$ has $S_z$=-1/2, it cannot form a transport channel
with $\alpha_T^{(+)}$ due to spin blockade.
Therefore, the transport channel $(\alpha_T^{(+)},\beta_{\uparrow})$
which carries spin-down current dominates the transport at high magnetic fields.
At higher bias, with $eV_{\rm sd}$=0.3 [Fig.~\ref{fig:QP2_spinAB_finiteV} (d)$\sim$(f)],
the pair of states $(\alpha_T^{(+)},\beta_{\uparrow})$ always forms a transport channel
and spin-down current is dominant at all magnetic fields.
The singlet state $\alpha_S$ can form transport channels with both $\beta_{\uparrow}$ and
$\beta_{\downarrow}$ states and contributes equally to the spin-up and spin-down currents.
The formation of transport channels involving spin-singlet $\alpha_S$ leads to
probability redistribution and decreases the probability of the triplet state $\alpha_T^{(+)}$.
Therefore, we have higher current polarization when the singlet state
$\alpha_S$ does not participate in the transport and vice versa.

\begin{figure}
  \includegraphics[width=0.8\linewidth]{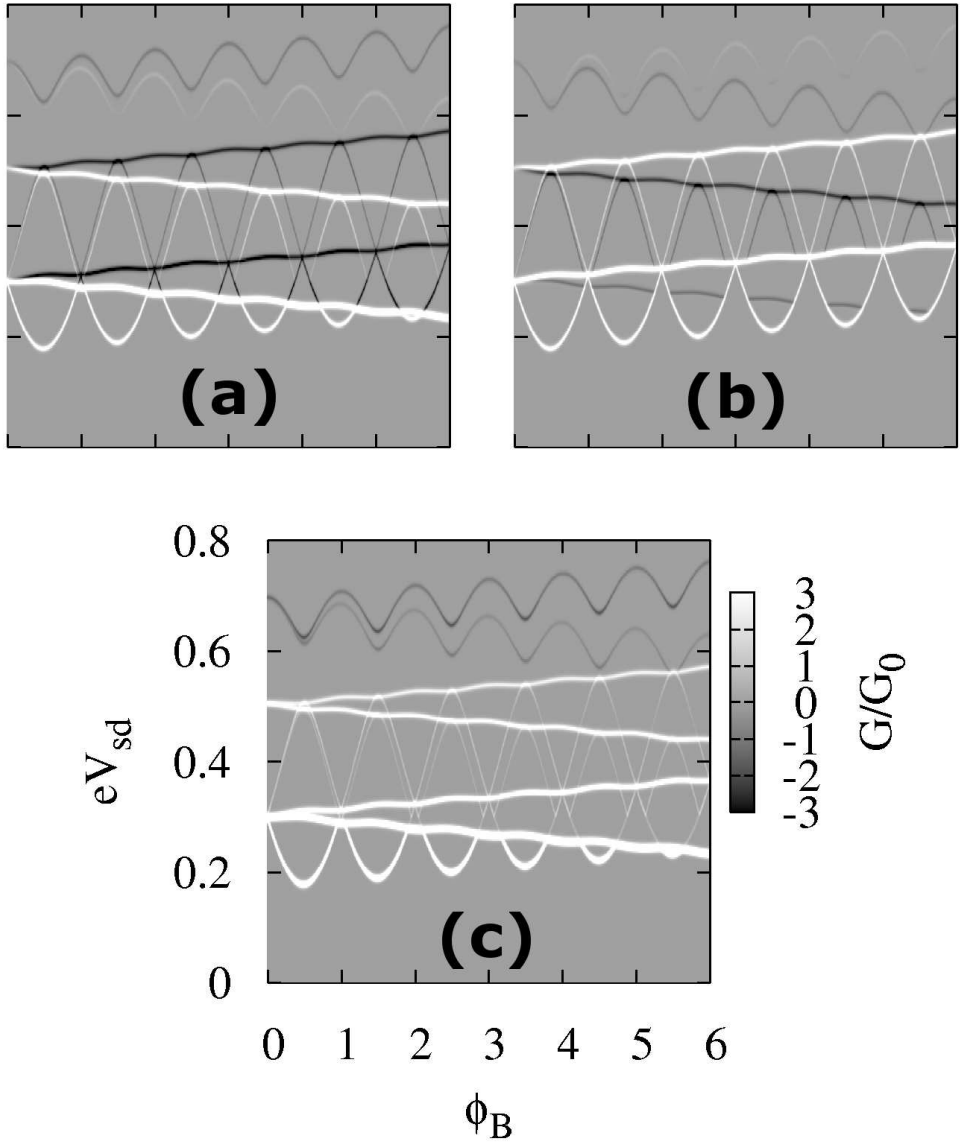}\\
  \caption{Differential conductance as a function of the magnetic field and the bias
           for the system in Fig.~\ref{fig:QP2_spectrum}(b).
           (a) and (b) show spin-down and spin-up components separately
           and (c) shows the total differential conductance. }
  \label{fig:QP2_DV_phi_G}
\end{figure}

Figure \ref{fig:QP2_DV_phi_G} summarizes both the negative differential conductance
and the spin-selective AB oscillations at finite bias for the system in Fig.~\ref{fig:QP2_spectrum}(b).
(a) and (b) show the spin-down and spin-up component of the differential conductance
as functions of magnetic field and the bias,
which clearly show different AB oscillations for spin-down and spin-up conductance.
(c) is the total differential conductance and the dark trace at large bias
due to introduction of transport channels involving the high energy singlet state
indicates the current decrease for increasing bias.
Since nonzero differential conductance indicates introduction of new transport channels
and the bias for new transport channel corresponds to
the energy difference between the two constituent states of the new transport channel,
the differential conductance can be used as a spectroscopic tool.
Comparing Figure \ref{fig:QP2_DV_phi_G}(c) with the energy spectrum Fig.~\ref{fig:QP2_spectrum}(b),
we can see the resemblance.
Differences are that singlet levels are split into two
and three Zeeman-split triplet levels give only two nonzero traces in differential conductance $G$.
The lowest singlet state forms transport channels with $\beta_{\uparrow}$ and $\beta_{\downarrow}$
at the same bias, as we mentioned earlier, and therefore leads to only one nonzero trace in $G$,
while higher singlet states can form transport channels with $\beta_{\uparrow}$ and $\beta_{\downarrow}$
at different bias and hence two nonzero traces in $G$.
For the triplet states, triplet states with $S_z$=+1 (-1)
can not form transport channel with $\beta_{\downarrow}$ ($\beta_{\uparrow}$) due to spin blockade
and therefore we have only two nonzero traces in $G$ for three triplet states.

\begin{figure}
  \includegraphics[width=0.8\linewidth]{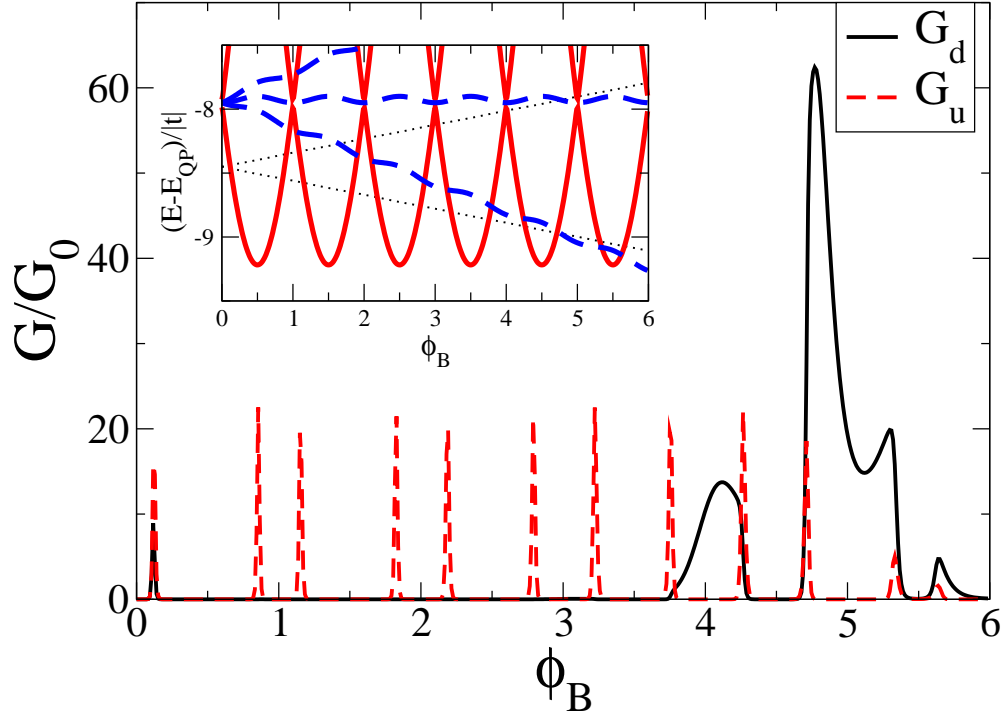}\\
  \caption{(Color online) Spin-selective AB oscillation around a QP
            with a trapped hole in dot 2 ($\vep_0=-0.025$).
            Spin-up conductance $G_u$ shows repeated peaks at lower magnetic fields
            and spin-down conductance $G_d$ has large peaks at large magnetic fields.
            Inset shows the energy spectrum which is shifted from the QP by $\vep_0=-0.025$.}
  \label{fig:QP2_spinAB}
\end{figure}

To compare with the spin-selective AB oscillations of TQD system with a trapped electron
in the linear response regime,\cite{delgado_shim_prl2008}
we consider the TQD slightly off the exact QP  with a trapped hole.
Figure~\ref{fig:QP2_spinAB} shows the spin-selective AB oscillations with $\vep_0=-0.025$
in the linear response regime where $eV_{\rm sd}=1.0\times 10^{-4}$.
The spin-up single-hole state crosses the singlet two-hole state several times
until it crosses with the triplet state at higher magnetic field.
The repeating conductance peaks are for spin-up current
and the large peak at strong enough magnetic field is for spin-down current.
Compared to the case at the QP with a trapped electron,
the behaviors of spin-up and spin-down currents are reversed
since the transport of a hole with spin-up (spin-down)
corresponds to the transport of an electron with spin-down (spin-up) in the opposite direction.
The phase of the oscillation is also shifted by $\pi$
due to the $\pi$ phase shift of oscillations of the singlet states.
The current is ultimately suppressed at very high field
where the triplet state is the ground state and
spin-up single-hole state is outside of the transport window.

\section{Conclusions}\label{sec:Conclusions}

We presented a theory of the tunneling transport through a TQD
around a QP with a trapped hole in dot 2 where spin-selective AB oscillations occur.
A detailed description of the formalism for the transport calculation
and general conditions for transport through multiple many-body states were given.
It was shown that the interplay between the introduction of new transport channels
and the probability redistribution can lead to negative differential conductance
and that the differential conductance can be used as a spectroscopic tool.
The transport in the magnetic field is sensitive to the spin of the carriers
and the spin structure of the TQD system in a triangular geometry
due to the strong Coulomb interaction and the interference effects.
This spin-selective AB oscillations show different behavior
in the linear response regime and with a finite bias.
The TQD system with singly connected leads can be considered as
a minimal quantum dot network (QDN) circuit and the formalism developed here
can be used for general QDN circuits.
The multi-channel transport can be analyzed using the spectral functions of the QDN
and transport channels formed by many-body states.

\begin{acknowledgments}
The authors thank M. Korkusinski, A. S. Sachrajda, S. A. Studenikin, and L. Gaudreau
for useful discussions. This work was supported
by the Canadian Institute for Advanced Research,
QuantumWorks, and NRC-CNRS CRP.
F. Delgado thanks support by the Juan de la Cierva program, Ministerio de Ciencia e Innovaci\'on.
\end{acknowledgments}


\begin{appendix}

\section{Noninteracting one-dimensional chain}\label{append:1Dchain}

The leads are described by the Hamiltonian of non-interacting one-dimensional chains
\begin{equation}
\nonumber
\widehat{H}_r = \sum_m\sum_{\sigma} \vep_0^r c^{\dag}_{m\sigma} c_{m\sigma}
               +\sum_m\sum_{\sigma} \left( t_r c^{\dag}_{m\sigma} c_{m+1\sigma} + h.c. \right)~,\label{eq:app.Hlead}
\end{equation}
where $\vep_0^r$ is the level energy of each site and $t_r$ is the tunneling between sites in lead $r=L,R$.
$m$ is from $-N_a$ to $-1$ for the left lead and from 1 to $N_a$ for the right lead.
The eigenstates of this tight-binding chain is
\begin{equation}
|k\sigma\rangle = \frac{1}{\sqrt{N_a}}\;\sum_{j=1}^{N_a} e^{ikaj}|j\sigma\rangle~,
\end{equation}
where $k=2n\pi/(N_a a)$ with $n=0,1,2,...,N_a-1$.
The eigenvalues are spin degenerate and given by
\begin{equation}
\varepsilon_{k}^r = \varepsilon_0^r + 2t_r \cos ka~,
\end{equation}
and the density of states per site for each spin is
\begin{equation}
\rho_{r\sigma}(\vep)
= \frac{1}{2\pi|t_r|\sin k_{\vep}a} \Theta\left(2 |t_r| - |\varepsilon-\varepsilon_0^r| \right)~,\label{eq:app.DOSlead}
\end{equation}
where $\Theta$ is the step function and $k_{\vep}$ is determined by $\vep=\vep_0^r+2t_r\cos(k_{\vep}a)$.

The TQD and the leads are connected by tunneling Hamiltonian
\beqa
\widehat{H}_{rD} &=& \sum_{\sigma} \left( t_{rD} c_{m_0\sigma}^{\dag} d_{i_0\sigma} + h.c \right)~,\label{eq:app.Hcouple}
\eeqa
where $m_0$ ($-1$ for $r=L$ and $1$ for $r=R$) and $i_0$ (1 for $r=L$ and 3 for $r=R$)
are the two adjacent sites of the lead and the dot connected by the tunneling
and $t_{rD}$ is the tunneling element connecting the two sites.
Using the eigenstates of the lead chains, we obtain
\begin{eqnarray}
\widehat{H}_{rD}
&=& \sum_{\sigma} \sum_k \left( \tilde{t}_{rD}(k) c^{\dag}_{k\sigma} d_{i_0\sigma} + h.c. \right)~,
\end{eqnarray}
where
\begin{equation}
\tilde{t}_{rD}(k) \equiv \frac{t_{rD} e^{-ikam_0}}{\sqrt{N_a}}~.\label{eq:app.t_tilde}
\end{equation}

\section{ Solution of the rate equation at the steady state }\label{append:steady_sol}

In this appendix, we present a method to find the steady state solution of the rate equation.
The master equation in matrix form is give by
\begin{equation}
\frac{d\mathbf{P}}{dt} = \mathbf{M}_P\cdot\mathbf{P}~,\label{eq:app.Master_matrix}
\end{equation}
where the matrix $\mathbf{M}_P$ is defined by Eq.~\eqref{eq:Master}.
The initial probabilities at $t=0$ are given by equilibrium values
\beqa
&& P_{\alpha_N}(0) = P_{\alpha_N}^{\rm eq}
 = \frac{ \exp\left( -\frac{ E_{\alpha_N}^D - \mu_0 N }{k_B T} \right)}{Z}~,\label{eq:app.P_eq}
\eeqa
For stationary cases ($t\rightarrow \infty$), $dP_{\alpha_N}/dt = 0$ for all $\alpha_N$.
Then we obtain a system of linear equations for $\alpha_N$
\begin{equation}
\mathbf{M}_P\cdot\mathbf{P} = 0~.\label{eq:app.Master_steady}
\end{equation}
This equation by itself does not uniquely determine $P_{\alpha_N}$
because the matrix $\mathbf{M}_P$ is singular.
We can see this by the fact that the summation of
all the elements of $\mathbf{M}_P$ is zero and therefore
the set of equations in Eq.~\eqref{eq:app.Master_steady} are not linearly independent.
We need more conditions to uniquely determine the steady state solution.
One constraint is the normalization condition of the probabilities:
\begin{equation}\label{eq:app.normalization_P}
\sum_N\sum_{\alpha_N}P_{\alpha_N}=1~.
\end{equation}
If all the states take part in the transport, then Eqs.~\eqref{eq:app.Master_steady} and ~\eqref{eq:app.normalization_P}
would uniquely determine the steady state probabilities.
But if some states don't participate in the transport, we need more conditions
(If we consider Eq.~\eqref{eq:app.Master_steady} as an eigenvalue problem with eigenvalue 0,
the eigenvalue 0 can be degenerate).
Considering that the master equation gives unique solution with the initial condition,
we need to make use of this initial condition to solve steady state solution in this case.

For this, let us use the singular value decomposition (SVD) of $\mathbf{M}_P$.
\begin{equation}
\mathbf{M}_P = \mathbf{U} \mathbf{D} \mathbf{V}^T~, \label{eq:app.SVD}
\end{equation}
where $\mathbf{U}$ and $\mathbf{V}$ are orthogonal matrices
and $\mathbf{D}$ is a diagonal matrix.
The diagonal elements of $\mathbf{D}$ are called {\it singular values}.
All singular values are zero or positive and we assume that the diagonal elements of $\mathbf{D}$
are in descending order.
The number of zero singular values ($N_0$) is the  dimension of the null space of $\mathbf{M}_P$,
and the last $N_0$ column vectors of $\mathbf{V}$ form a basis set for the null space.
If only one singular value is zero,
the last column vector of $\mathbf{V}$ defines the one-dimensional null space of $\mathbf{M}_P$
and the the normalization condition would be enough to uniquely determine the probabilities.
But, in general, there can be more than one 0 singular values.
For $N_P\times N_P$ matrix $\mathbf{M}_P$, $N_P=N_R+N_0$ where $N_R$ is the dimension of the range
and $N_0$ is the dimension of the null space.
Eq.~\eqref{eq:app.Master_matrix} can be written as
\begin{eqnarray}
\frac{d\mathbf{z}}{dt} &=& \mathbf{D}\mathbf{y}~,
\end{eqnarray}
where
\begin{eqnarray}
\mathbf{z} \equiv \mathbf{U}^T \mathbf{P}~, \quad
\mathbf{y} \equiv \mathbf{V}^T \mathbf{P}~.
\end{eqnarray}
Separating the null space ($B$) and its complementary space ($A$), it becomes
\begin{eqnarray}
&& \frac{d}{dt} \left( \begin{array}{c} \mathbf{z}_A \\ \mathbf{z}_B \end{array} \right)
= \left( \begin{array}{cc} \mathbf{D}_A & 0 \\ 0 & 0 \end{array} \right)
  \left( \begin{array}{c} \mathbf{y}_A \\ \mathbf{y}_B \end{array} \right)  \nonumber\\
&\Rightarrow& \frac{d\mathbf{z}_A}{dt} = \mathbf{D}_A \mathbf{y}_A ~, \quad \frac{d\mathbf{z}_B}{dt} = 0~.
\end{eqnarray}
$\mathbf{D}_A$ is a diagonal matrix whose diagonal elements are the nonzero singular values of $\mathbf{M}_P$.
Since time derivative of $\mathbf{z}_B$ is zero, we obtain
\begin{equation}
\mathbf{z}_B (t) = \mathbf{z}_B (0)
\end{equation}
at all time $t$.
At $t\rightarrow \infty$, all time derivatives are zero and
we obtain
\begin{eqnarray}
&& \frac{d\mathbf{z}_A(t\rightarrow \infty)}{dt} = \mathbf{D}_A \mathbf{y}_A (t\rightarrow \infty) = 0 \\
\Rightarrow && \mathbf{y}_A (t\rightarrow \infty) = 0
\end{eqnarray}
since $\mathbf{D}_A$ is a diagonal matrix with all nonzero diagonal elements.
Using
\begin{eqnarray}
&& \mathbf{y} = \mathbf{V}^T \mathbf{P} = \mathbf{V}^T \mathbf{U} \mathbf{z} \equiv \mathbf{W} \mathbf{z} ~, \\
\Rightarrow&&
\left( \begin{array}{c} \mathbf{y}_A \\ \mathbf{y}_B \end{array} \right)
= \left( \begin{array}{cc} \mathbf{W}_{AA} & \mathbf{W}_{AB} \\ \mathbf{W}_{BA} & \mathbf{W}_{BB} \end{array} \right)
\left( \begin{array}{c} \mathbf{z}_A \\ \mathbf{z}_B \end{array} \right)~,
\end{eqnarray}
we get
\begin{eqnarray}
\mathbf{y}_A (\infty) = \mathbf{W}_{AA}\mathbf{z}_A(\infty) + \mathbf{W}_{AB} \mathbf{z}_B(\infty) = 0
\end{eqnarray}
and we obtain an equation for $\mathbf{z}_A(\infty)$
\begin{eqnarray}
&& \mathbf{W}_{AA}\mathbf{z}_A(\infty) = - \mathbf{W}_{AB} \mathbf{z}_B(\infty) \\
\Rightarrow && \mathbf{z}_A(\infty) = -\left(\mathbf{W}_{AA}\right)^{-1} \mathbf{W}_{AB} \mathbf{z}_B(0)~,
\end{eqnarray}
where we used the result that $\mathbf{z}_B$ is constant in time.
We can prove that $\left(\mathbf{W}_{AA}\right)^{-1}$ exists as follows.
Since $\mathbf{M}_P$ has a null space of dimension $N_0$,
$\mathbf{M}_P$ has eigenvalue 0 of degeneracy $N_0$.
From Eq.~\eqref{eq:app.SVD}, we have
\beq\label{eq:app.det}
\mathbf{U}^T \mathbf{M}_P \mathbf{U} = \mathbf{D} \mathbf{W}
= \left(\begin{array}{cc}
        \mathbf{D}_A \mathbf{W}_{AA} & \mathbf{D}_A \mathbf{W}_{AB} \\ 0 & 0
        \end{array}
  \right)~.
\eeq
Since $\mathbf{U}^T \mathbf{M}_P \mathbf{U}$ has also eigenvalue 0 of degeneracy $N_0$,
\beq
{\rm det} \left( \mathbf{U}^T \mathbf{M}_P \mathbf{U} - \lambda \mathbf{I} \right)
= \lambda^{N_0} f(\lambda) ~,
\eeq
where $f(\lambda)$ is a polynomial with $f(0)\neq 0$.
Using Eq.~\eqref{eq:app.det},
\beqa
{\rm det}\left( \mathbf{D}\mathbf{W} -\lambda \mathbf{I} \right)
&=& \lambda^{N_0} {\rm det} \left( \mathbf{D}_A \mathbf{W}_{AA} -\lambda \mathbf{I}_{AA} \right)
\eeqa
and $\lambda=0$ is not a solution of
${\rm det} \left( \mathbf{D}_A \mathbf{W}_{AA} - \lambda \mathbf{I}_{AA} \right) = 0$.
Therefore,
\beq
{\rm det} \left( \mathbf{D}_A \mathbf{W}_{AA} \right)
= {\rm det} \mathbf{D}_A \, {\rm det}\mathbf{W}_{AA} \neq 0~,
\eeq
and, since $\mathbf{D}_A$ is a diagonal matrix with nonzero elements, we obtain
${\rm det}\mathbf{W}_{AA} \neq 0$ and $\left(\mathbf{W}_{AA}\right)^{-1}$ exists.
Once we have $\mathbf{z}(\infty)$, we can find $\mathbf{P}$ at steady state using
\begin{equation}
\mathbf{P}(t\rightarrow \infty) = \mathbf{U} \mathbf{z}(\infty)~.
\end{equation}
The normalization condition is automatically satisfied in this method,
because the master equation conserves the normalization.

\end{appendix}




\newpage



\end{document}